\documentclass[conference]{IEEEtran}
\IEEEoverridecommandlockouts
\usepackage{cite}
\usepackage{amsmath,amssymb,amsfonts}
\usepackage{algorithmic}
\usepackage{graphicx}
\usepackage{textcomp}
\usepackage{xcolor}

\def\BibTeX{{\rm B\kern-.05em{\sc i\kern-.025em b}\kern-.08em
    T\kern-.1667em\lower.7ex\hbox{E}\kern-.125emX}}
\begin{document}

\title{
Analog-Memory-Based Waveform Generation for Qubit Control
\thanks{
This work was supported by JST Moonshot R\&D (Grant Number JPMJMS256G and  JPMJMS256L).
}
}

\author{
\IEEEauthorblockN{
Shinichi~Morisaka\IEEEauthorrefmark{1}\IEEEauthorrefmark{2},
Toshiaki~Inada\IEEEauthorrefmark{3},
Yoshinori Kurimoto\IEEEauthorrefmark{1},
Atsushi~Noguchi\IEEEauthorrefmark{4}\IEEEauthorrefmark{5}\IEEEauthorrefmark{6},
and Ryutaro~Ohira\IEEEauthorrefmark{1}
}

\IEEEauthorblockA{
\IEEEauthorrefmark{1}QuEL, Inc., Hachioji ON Building 5F, 4-7-14 Myojincho, Hachioji, Tokyo, Japan
}
\IEEEauthorblockA{
\IEEEauthorrefmark{2}Center for Quantum Information and Quantum Biology, \\The University of Osaka, 1-2 Machikaneyama, Toyonaka, Osaka, Japan
}
\IEEEauthorblockA{
\IEEEauthorrefmark{3}International Center for Elementary Particle Physics (ICEPP), \\The University of Tokyo, 7-3-1 Hongo, Bunkyo, 113-0033, Tokyo, Japan
}
\IEEEauthorblockA{
\IEEEauthorrefmark{4}Komaba Institute for Science (KIS), The University of Tokyo, 3-8-1 Komaba, Meguro, 153-8902, Tokyo, Japan
}
\IEEEauthorblockA{
\IEEEauthorrefmark{5}RIKEN Center for Quantum Computing (RQC), 2-1 Hirosawa, Wako, 351-0198, Saitama, Japan
}
\IEEEauthorblockA{
\IEEEauthorrefmark{6}Inamori Research Institute for Science (InaRIS), 620 Suiginya, Kyoto, 600-8411, Kyoto, Japan
}

\IEEEauthorblockA{
Emails: morisaka@quel-inc.com, tinada@icepp.s.u-tokyo.ac.jp, kurimoto@quel-inc.com, \\u-atsushi@g.ecc.u-tokyo.ac.jp, ohira@quel-inc.com
}
}

\maketitle

\begin{abstract}
We propose an analog-memory-based arbitrary waveform generator for cryogenic qubit control that separates low-speed waveform programming from high-speed waveform generation.
Waveform samples are stored as capacitor voltages and sequentially accessed through buffered switching networks, enabling repeated, non-destructive readout without high-speed digital-to-analog conversion.
A cryogenic prototype demonstrated Gaussian-shaped waveform generation at 14~K with a sampling rate of 40~Msps using a programming rate of 50~ksps.
No measurable voltage decay was observed over 2~s at cryogenic temperatures.
A complementary room-temperature prototype achieved 160~Msps.
The architecture offers a potential route toward lower-power cryogenic waveform generation.
\end{abstract}


\section{Introduction}
\label{sec:introduction}

The scaling of quantum processors places increasing demands on the classical electronics required for qubit control~\cite{krantz2019quantum, rizvi2026survey}.
In conventional superconducting quantum computing systems, shaped microwave pulses are generated at room temperature and delivered to qubits through cables entering the cryostat~\cite{xu2021qubic, stefanazzi2022qick, park2022icarus, machino2026qube, kurimoto2026microwave}.
As the number of qubits increases, the associated cable count and conductive heat load become important system-level constraints.
Integrating waveform-generation circuits at cryogenic temperatures is therefore a promising approach to reducing the complexity of the quantum--classical interface~\cite{hornibrook2015cryogenic, charbon2016cryo}.

Cryogenic integration can reduce the number of signals routed from room temperature, but it also requires waveform-generation hardware to operate within a limited cooling-power budget. 
In a conventional arbitrary waveform generator (AWG), high-rate digital sample delivery and digital-to-analog conversion take place whenever a programmed waveform is produced. 
These operations are performed repeatedly even when the same calibrated waveform is reused for many qubit gates. 
This mismatch between infrequent waveform updates and fast pulse generation motivates a different organization of waveform storage and readout.

In this work, we propose an analog-memory-based AWG that decouples waveform programming from high-speed waveform generation.
Waveform samples are stored as voltages in capacitor-based memory cells and sequentially accessed through a high-speed switching network.
By isolating the stored voltages from the readout path, the proposed architecture enables non-destructive, repeated waveform generation without requiring high-speed digital-to-analog conversion.
We demonstrate Gaussian-shaped pulse generation at 14~K with a sampling rate of 40~mega-samples per second (Msps) using a programming rate
of 50~kilo-samples per second (ksps).
A complementary room-temperature prototype achieves a sampling rate of 160~Msps.
These results establish the feasibility of generating high-speed waveforms from locally stored analog samples under cryogenic conditions.
\section{Background and Design Motivation}
\label{sec:background}

\subsection{Conventional Arbitrary Waveform Generators}
\label{subsec:conventional_awg}

Conventional AWGs store waveform samples digitally and sequentially convert them into analog signals using high-speed digital-to-analog converters (DACs).
This architecture is widely used for generating shaped pulses in superconducting quantum computing systems~\cite{xu2021qubic, stefanazzi2022qick, park2022icarus, machino2026qube, kurimoto2026microwave}.

Cryogenic waveform generators have also been developed using CMOS technologies~\cite{xue2021cmos, park2021fully, prathapan2022cryogenic, underwood2024using}.
Many existing implementations rely on digital waveform storage or control data and high-speed digital-to-analog conversion during pulse generation.
Alternatively, direct-synthesis circuits can generate specific pulse shapes without waveform memory~\cite{kang202334}.
These approaches motivate alternative architectures that retain arbitrary waveform programmability while reducing the need for high-speed digital-to-analog conversion.

In conventional DAC-based architectures, waveform amplitudes are converted from digital values into analog voltages at the waveform sampling rate.
Separating waveform-amplitude conversion from high-speed waveform generation therefore offers the potential to reduce power consumption.
However, the overall power consumption also depends on the circuitry required for waveform readout, switching control, and output buffering.

\subsection{Analog Memory and Related Architectures}
\label{subsec:analog_memory_architecture}

Analog voltage storage has been investigated for cryogenic quantum-control applications, including semiconductor-qubit gate control~\cite{xu2020chip, pauka2021cryogenic, ferraris2024cryogenic, bartee2025spin} and trapped-ion electrode control~\cite{malinowski2023wire, ohira2025multiplexed, ohira2025trapping, ohira2026cryogenic, anmasser2026demonstration}.
These architectures primarily retain static or slowly varying control voltages rather than temporal sample sequences for high-speed waveform generation.

Analog sample storage and sequential access have also been investigated in bucket-brigade circuits, analog waveform generators, and switched-capacitor waveform recorders~\cite{sangster1969bucket, brunn1975analog,
chawla2005analog, kleinfelder19904096, ritt2008design}.
These studies establish the use of analog memory for storing and accessing waveform samples in various applications.

In particular, analog waveform recorders typically acquire transient signals at high speed and subsequently read out the stored samples for measurement.
In contrast, the proposed architecture programs a prescribed sequence of analog voltages and repeatedly accesses the stored samples to generate control pulses.
Unlike sample-and-hold voltage distribution, which maintains approximately constant voltages at individual electrodes or gates, the proposed approach uses sequential readout of analog memory cells to reconstruct time-dependent waveforms.

The proposed architecture differs from conventional analog waveform recorders in its intended operation.
Rather than capturing transient signals for subsequent measurement, it programs prescribed waveform samples and repeatedly generates the stored waveforms through buffered, non-destructive readout.
Combined with a shared low-speed programming source, this organization allows multiple output channels to reuse calibrated waveforms without high-speed digital-to-analog conversion during pulse generation.

\subsection{Design Motivation}
\label{subsec:design_motivation}

Qubit-control waveforms are typically reused across many gate operations after calibration.
Although pulse generation requires high sampling rates, the waveform samples need to be updated only when the pulse shape or calibration parameters change.
This difference in timescales motivates separating waveform programming from high-speed waveform generation.

In the proposed approach, waveform samples are stored as voltages in capacitor-based analog memory cells.
During pulse generation, the stored voltages are sequentially accessed through a high-speed switching network.
The memory nodes are isolated from the readout path, allowing the stored voltages to be accessed without directly discharging the storage capacitors.
This non-destructive readout enables repeated waveform generation without high-speed digital-to-analog conversion during pulse generation.

In practice, voltage retention and waveform accuracy are affected by leakage currents, parasitic coupling, and switching-induced disturbances.
Nevertheless, separating waveform programming from high-speed readout allows the stored samples to be reused as long as their voltage errors remain within the required tolerance.

Another potential advantage of the proposed architecture is reduced switching activity during waveform generation.
In conventional high-speed DACs, digital input codes control multiple internal switching elements to generate the corresponding analog output levels.
Depending on the DAC architecture and the input code transition, multiple switches may change their states within a single sampling period, contributing to dynamic power consumption.

In contrast, the proposed architecture stores the waveform samples as analog voltages and generates waveforms by sequentially selecting the stored values through a readout multiplexer.
In a one-hot switching configuration, only one memory cell is selected at a time, and each transition between successive samples requires turning off the previously selected switch and turning on the next switch.
Thus, the number of switching elements that change states at each sample transition remains constant, regardless of the number of stored waveform samples.
This limited switching activity may reduce the dynamic power consumption associated with waveform generation, although the total power consumption also depends on the readout buffers and digital control circuitry.

\section{Proposed Analog-Memory-Based AWG Architecture}
\label{sec:architecture}

\begin{figure}[t]
\centering
\includegraphics[width=\columnwidth]{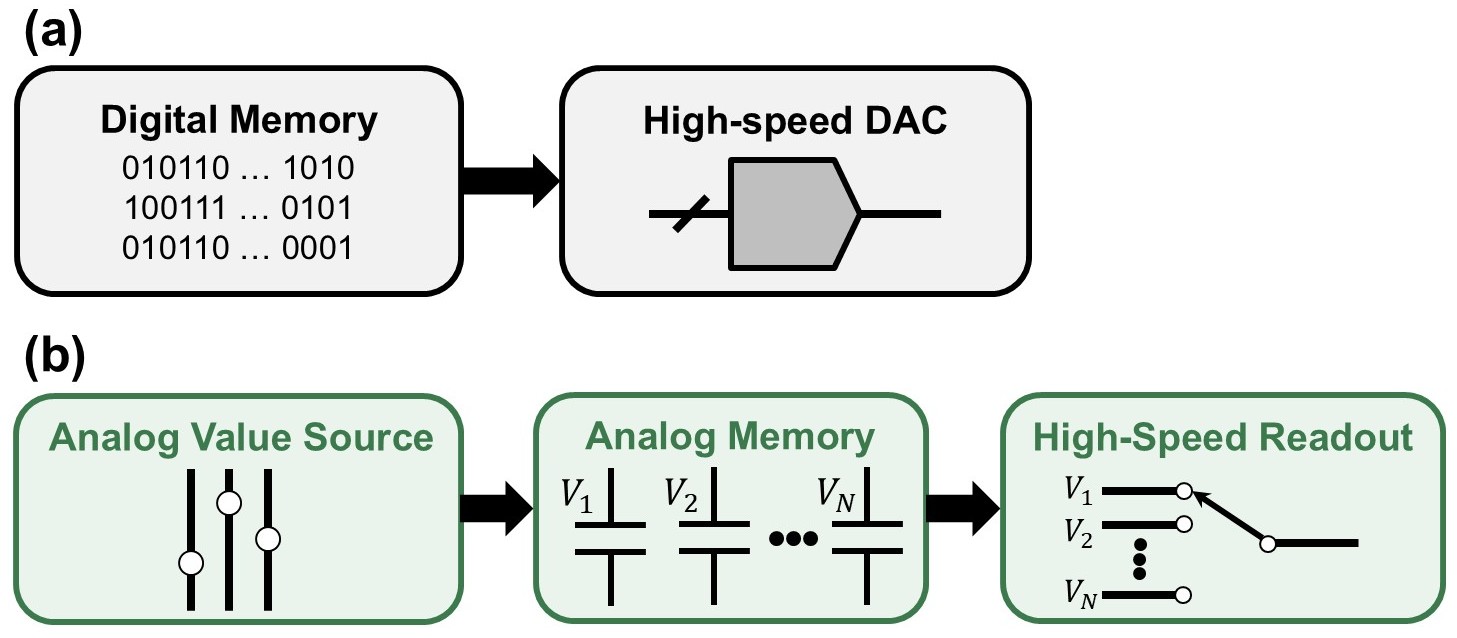}
\caption{
Comparison of waveform-generation architectures.
(a) Conventional AWG using digital waveform memory and a high-speed DAC.
(b) Proposed AWG using analog waveform memory and a high-speed switching network.
}
\label{fig:concept}
\end{figure}

\subsection{System Architecture}
\label{subsec:system_architecture}

Fig.~\ref{fig:concept} compares conventional and proposed AWG architectures.
In conventional AWGs, digitally stored waveform samples are converted into analog signals using a high-speed DAC during waveform generation.
In contrast, the proposed architecture stores waveform samples as analog voltages and sequentially accesses them through a high-speed switching network, eliminating high-speed digital-to-analog conversion from the waveform-generation path.

\begin{figure}[t]
\centering
\includegraphics[width=\columnwidth]{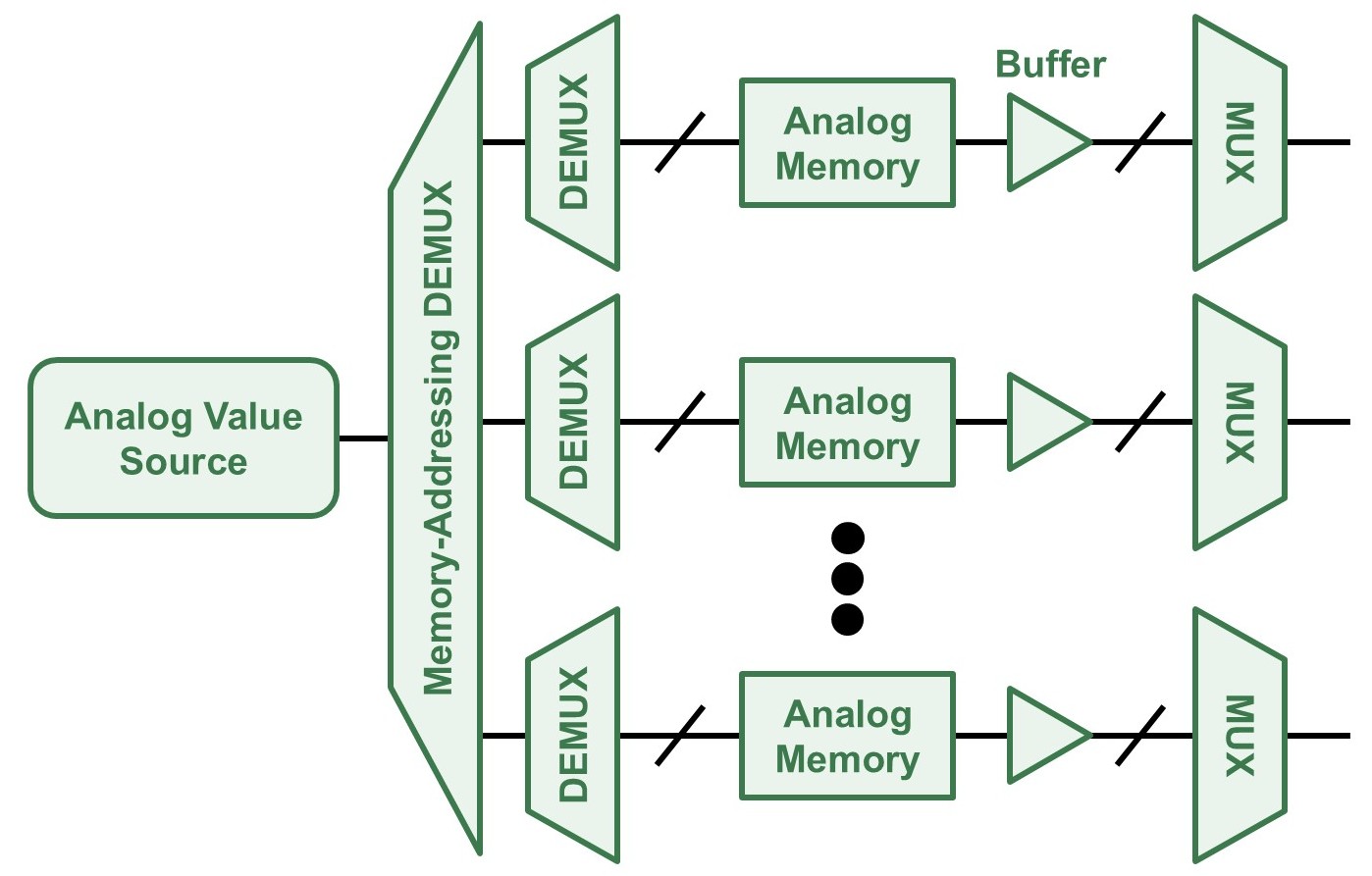}
\caption{
Multi-channel analog-memory-based AWG architecture.
A shared analog voltage source programs multiple memory arrays through cascaded demultiplexers.
Readout buffers and high-speed multiplexers generate independent waveforms from the stored samples.
}
\label{fig:architecture}
\end{figure}

Fig.~\ref{fig:architecture} illustrates the proposed multi-channel architecture.
A shared analog voltage source programs multiple analog memory arrays through cascaded demultiplexers.
Each array stores a sequence of voltage samples corresponding to an output waveform.
During waveform generation, the stored voltages are sensed by readout buffers and sequentially selected through high-speed multiplexers.
This configuration allows multiple channels to generate independent waveforms while sharing the programming source.

A waveform is represented by $N$ voltage samples, $\{V_1,V_2,\ldots,V_N\}$.
During programming, these values are sequentially written into the selected memory cells.
During waveform generation, the readout multiplexer sequentially selects the stored samples at a sampling rate of $f_{\mathrm{s}}$, producing a multilevel sample-and-hold waveform.
The programming rate and waveform-generation rate are independently controlled.

\begin{figure*}[t]
\centering
\includegraphics[width=\textwidth]{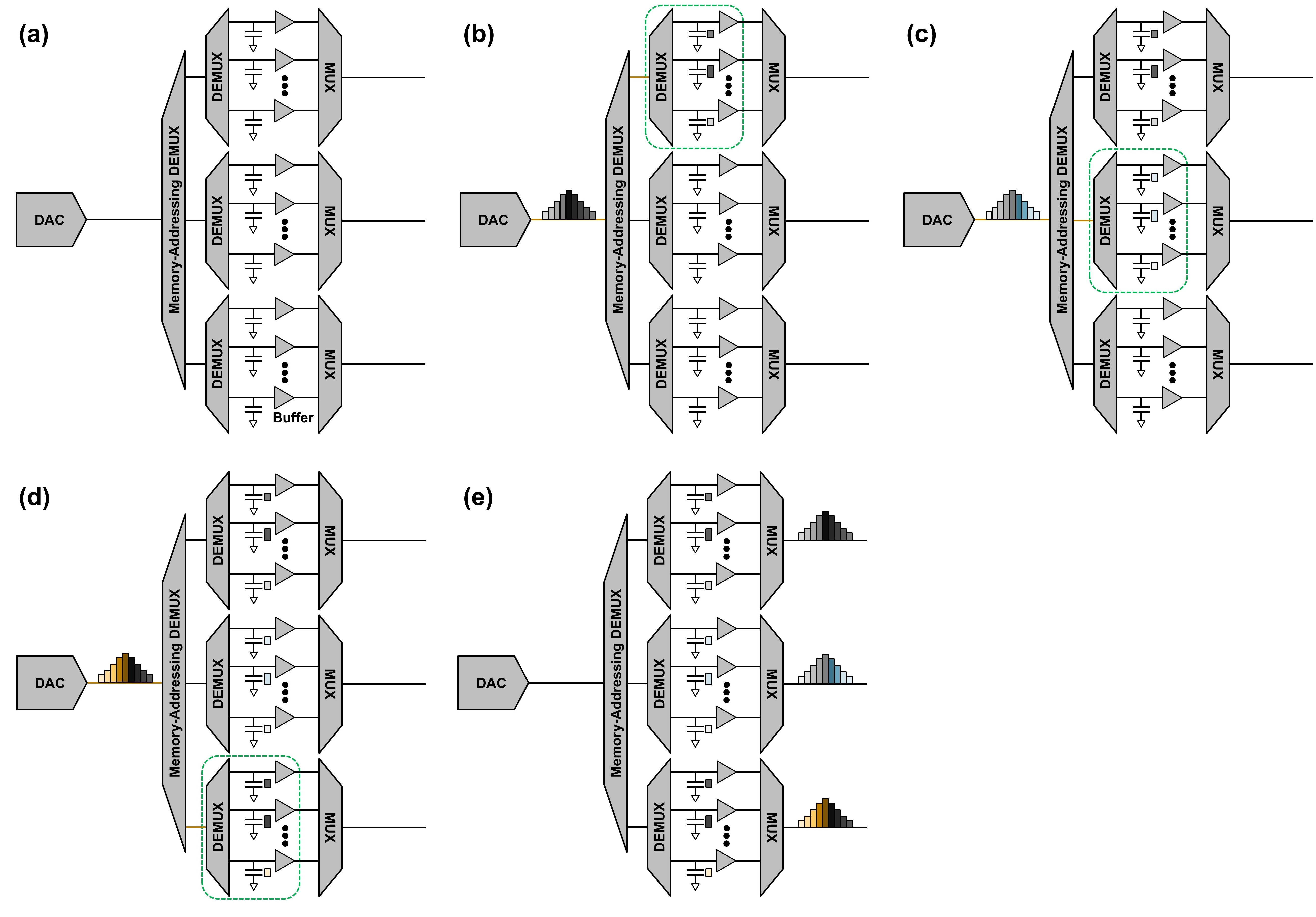}
\caption{
Operating sequence of the proposed multi-channel analog-memory-based AWG.
(a) Initial state.
(b)--(d) Sequential programming of different waveforms into selected analog memory arrays using a shared programming source.
(e) Independent waveform generation from the programmed memory arrays.
}
\label{fig:implementation}
\end{figure*}

The readout buffers isolate the memory cells from the switching network, allowing the stored voltages to be accessed without directly discharging the storage capacitors.
This non-destructive readout enables repeated waveform generation without reprogramming, provided that leakage and readout-induced disturbances remain within the allowable voltage error.

\subsection{Multi-Channel Programming and Waveform Generation}
\label{subsec:illustrative_implementation}

Fig.~\ref{fig:implementation} illustrates an example implementation of the proposed architecture and its operating sequence.
Fig.~\ref{fig:implementation}(a) shows the initial state, in which the analog memory arrays have not yet been programmed.
During programming, a shared DAC supplies the voltage samples of each waveform, while a hierarchical demultiplexer network selects the target memory array and individual memory cells.
As shown in Figs.~\ref{fig:implementation}(b)--(d), different waveform sequences are sequentially written into the selected memory arrays, with each cell storing the voltage corresponding to one waveform sample.

Once programming is complete, each memory array can independently generate its stored waveform, as illustrated in Fig.~\ref{fig:implementation}(e).
The readout multiplexers sequentially select the buffered voltages from the memory cells, producing independent output waveforms from the programmed arrays.
Since waveform generation does not require operation of the shared programming DAC, multiple channels can generate different waveforms simultaneously.

Since waveform programming and generation are performed independently, the programming rate can be much lower than the waveform-generation rate.
Once programmed, the stored waveforms can be generated repeatedly without updating the memory cells.
This architecture is therefore particularly suitable for qubit-control applications in which calibrated pulses are reused across multiple gate operations.

\subsection{Hybrid Architecture for Qubit Calibration}
\label{subsec:hybrid_architecture}

Although the proposed architecture eliminates high-speed digital-to-analog conversion during waveform generation, updating stored waveforms requires sequential programming of analog memory cells.
This process can introduce additional latency during qubit calibration, when pulse parameters are frequently adjusted.

\begin{figure}[t]
\centering
\includegraphics[width=\columnwidth]{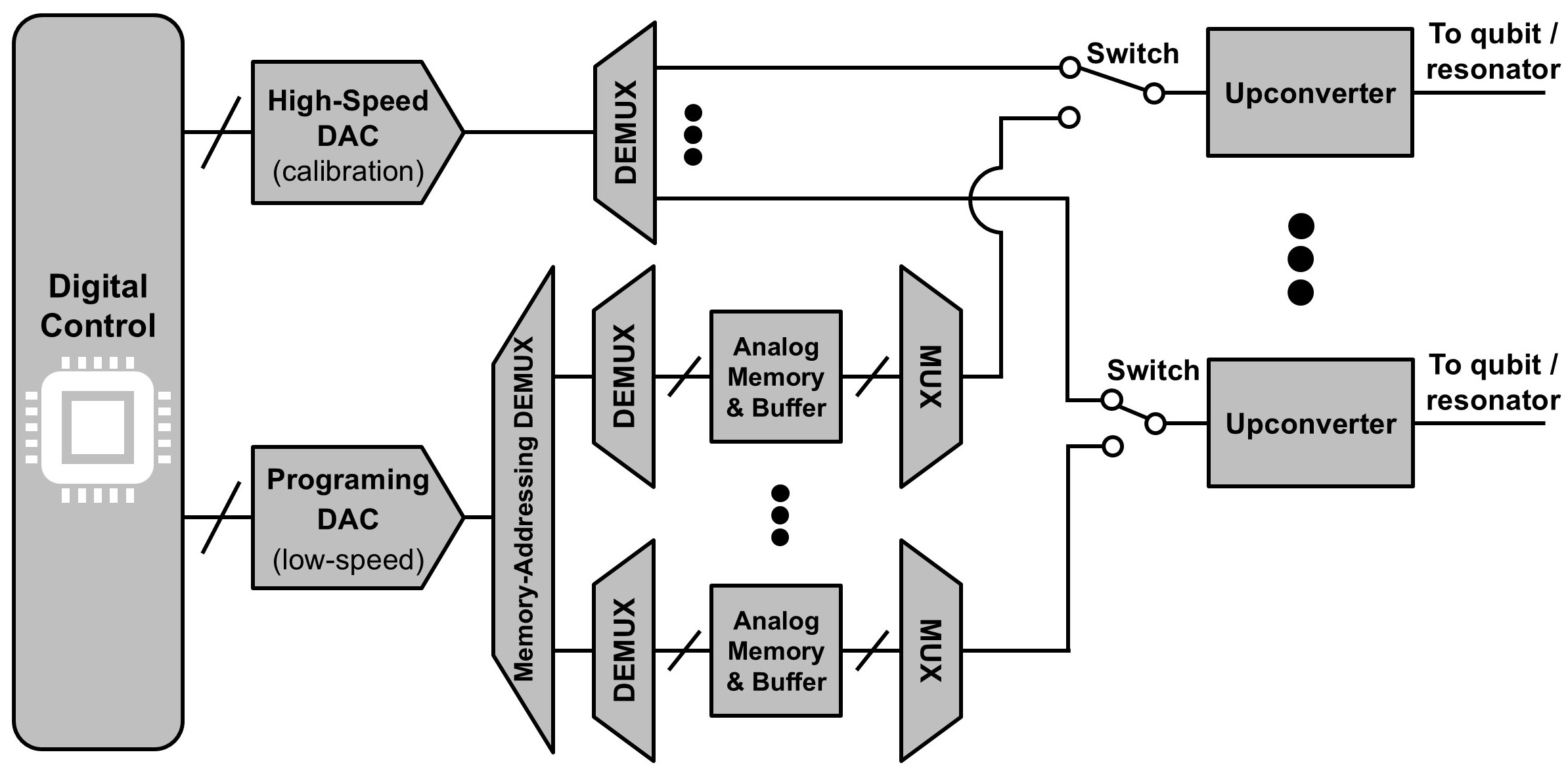}
\caption{
Conceptual hybrid quantum-control architecture combining a shared high-speed DAC for qubit calibration with analog-memory-based AWGs for repeated waveform generation.
Optimized digital waveform samples are programmed into the analog memory arrays through a shared programming DAC.
Each channel selects either the high-speed DAC or analog-memory output for subsequent upconversion.
}
\label{fig:hybrid}
\end{figure}

To address this limitation, we propose a hybrid architecture combining analog-memory-based AWGs with a shared high-speed DAC, as illustrated in Fig.~\ref{fig:hybrid}.
During calibration, the high-speed DAC generates programmable control pulses for a selected channel, and the pulse parameters are optimized based on the measured qubit response.
Once calibration is complete, the optimized digital waveform samples are transferred to the corresponding analog memory array through a shared programming DAC and a hierarchical demultiplexer network.

During subsequent quantum circuit execution, the stored waveforms are generated through the analog-memory paths without high-speed digital-to-analog conversion.
Each channel includes a switch that selects either the high-speed DAC output or the analog-memory output, allowing both paths to share the same upconversion circuitry.
The high-speed DAC can therefore be shared across multiple channels for calibration, while the analog-memory arrays provide independent waveform generation during quantum circuit execution.

The hybrid architecture is proposed as a system-level extension and has not been experimentally demonstrated in this work.
Since the high-speed DAC and analog-memory paths may exhibit different output characteristics, additional calibration may be required after transferring the waveform data to analog memory.

\section{Experimental Demonstration}
\label{sec:experimental}

\subsection{Cryogenic Prototype and Characterization}
\label{subsec:cryogenic}

\begin{figure}[t]
\centering
\includegraphics[width=\columnwidth]{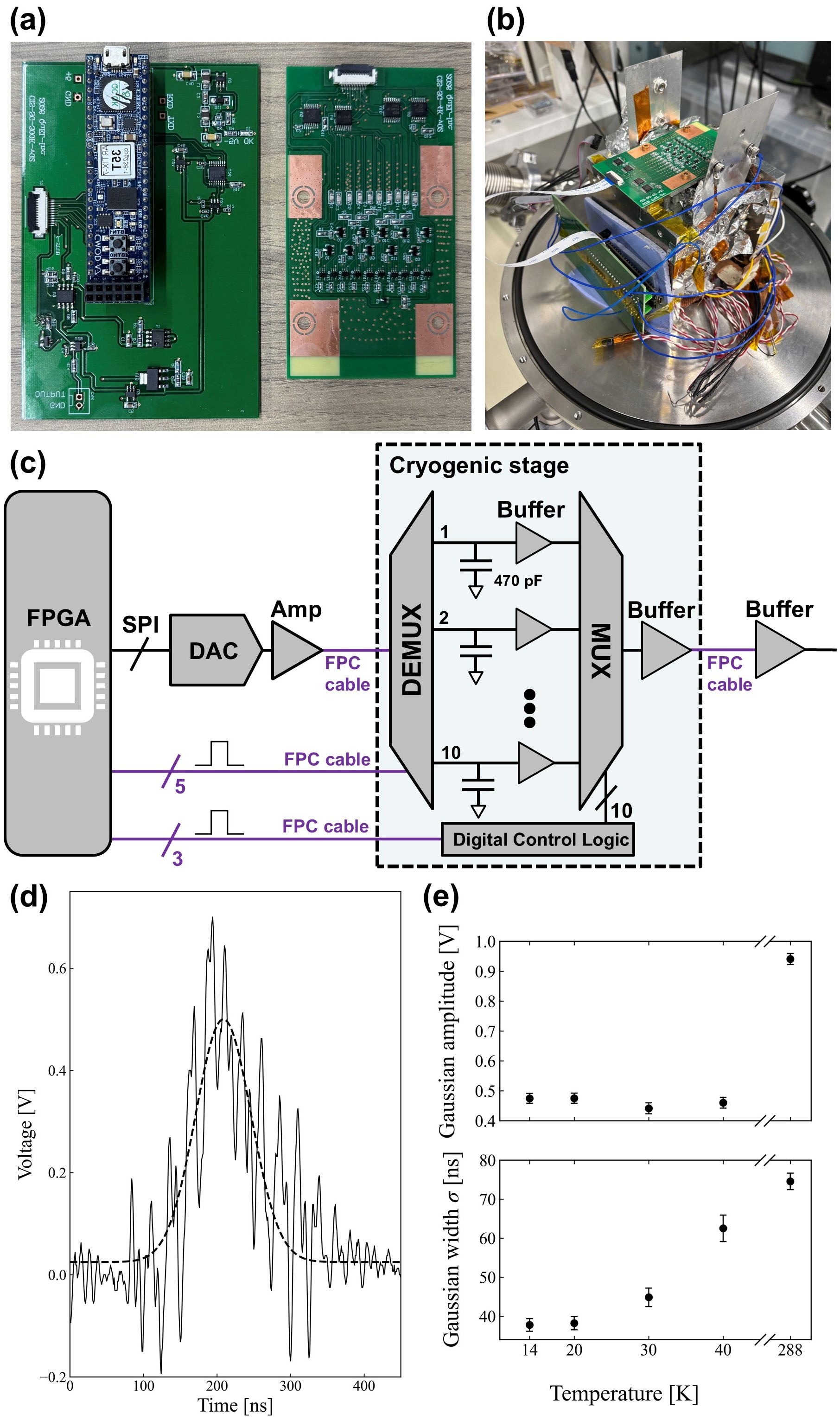}
\caption{
Cryogenic prototype and experimental results.
(a) Developed circuit boards.
(b) Experimental setup.
(c) Simplified circuit configuration.
(d) Gaussian-shaped waveform generated at 14~K with a sampling rate of 40~Msps.
The dashed curve is a Gaussian fit to guide the eye.
(e) Temperature dependence of the Gaussian amplitude (top) and width (bottom), extracted from Gaussian fits to the measured waveforms. 
Error bars represent one standard error obtained from the fitting covariance matrices. 
The horizontal axis is broken between 40~K and 288~K.
}
\label{fig:cryogenic}
\end{figure}

A cryogenic prototype was developed to evaluate analog-memory-based waveform generation at low temperatures.
Figs.~\ref{fig:cryogenic}(a)--(c) show the developed circuit boards, cryogenic experimental setup, and simplified circuit configuration, respectively.

As shown in Fig.~\ref{fig:cryogenic}(a), the system consists of two circuit boards: a room-temperature control board and a cryogenic board containing the analog memory and readout circuitry.

The room-temperature control board employs a 12-bit DAC (AD5328, Analog Devices) controlled by a Digilent Cmod A7 FPGA board through a serial peripheral interface (SPI).
Waveform samples are programmed at a rate of 50~ksps.
The DAC output is amplified with a gain of two using an operational amplifier (TLV271CW5, Texas Instruments).
The FPGA also generates the digital signals required to operate the demultiplexer and digital control logic on the cryogenic board.
The amplified analog signal and digital control signals are transmitted to the cryogenic board through a flexible printed circuit (FPC) cable.

The cryogenic board contains ten analog memory cells, each incorporating a 470-pF C0G capacitor for voltage storage.
The programming voltages are distributed to the memory cells through a demultiplexer network comprising two TMUX6208 devices (Texas Instruments).
During waveform generation, the stored voltages are sensed by MOSFET-based source followers (T2N7002AK, Toshiba Semiconductor and Storage) and sequentially selected through a readout multiplexer implemented using analog switches (SN74AUC1G66, Texas Instruments).
The readout sequence is controlled by digital logic incorporating two 8-bit shift registers (SN74LV164A, Texas Instruments).

The source followers provide high-input-impedance voltage sensing, electrically isolating the storage capacitors from the readout switching
network.
Consequently, waveform generation does not require transferring or intentionally discharging the stored charge, allowing the same analog
samples to be accessed repeatedly.
The generated signal is transmitted back to the room-temperature board through the FPC cable and delivered to an output buffer (LT1818CS5, Analog Devices).
The buffered output is measured using an oscilloscope.

Fig.~\ref{fig:cryogenic}(b) shows the cryogenic experimental setup.
Measurements were performed using an RDC-02K cryocooler (Sumitomo Heavy Industries, Ltd.).
The minimum measured board temperature was 14~K.
The cryocooler has specified cooling capacities of 1.0~W at 60~K for the first stage and 0.02~W at 2.3~K for the second stage.
The temperature was measured using a sensor mounted on the top surface of the circuit board under evaluation and therefore does not necessarily represent the cold-head temperature.
Furthermore, owing to board-size constraints, the demultiplexer and digital control circuitry were positioned away from the cold head without direct thermal contact.
Consequently, their operating temperatures may have been higher than the measured 14~K.

Fig.~\ref{fig:cryogenic}(d) shows a Gaussian-shaped waveform generated at 14~K with a sampling rate of 40~Msps.
The dashed curve represents a Gaussian fit included as a guide to the eye.
The measured waveform follows the programmed Gaussian envelope, demonstrating waveform generation from analog samples stored at
cryogenic temperatures.

However, periodic voltage spikes are superimposed on the generated waveform.
These spikes are likely caused by crosstalk from the 40-MHz digital switching signals into the analog signal path.
In the present implementation, the digital control signals and analog output are transmitted through a fine-pitch FPC cable, resulting in close proximity between the digital and analog signal lines.
The observed interference is therefore considered a signal-routing issue rather than an intrinsic limitation of the analog-memory-based architecture.
Separating the digital and analog signal paths using dedicated cables or
lower-crosstalk interconnects is expected to mitigate this interference.

Fig.~\ref{fig:cryogenic}(e) shows the temperature dependence of the amplitude and width (standard deviation $\sigma$) extracted from Gaussian fits to the measured waveforms.
The fitted amplitude decreases from approximately 0.94~V at 288~K to 0.48~V at 14~K, with relatively small variations between 14~K and 40~K. 
The Gaussian width also decreases with cooling, from approximately 75~ns at 288~K to 38~ns at 14~K, with little variation between 14~K and 20~K. 
These results indicate temperature-dependent changes in both the amplitude and temporal profile of the generated waveforms. 
The observed variations may be associated with the temperature-dependent characteristics of the MOSFET source followers and analog switches, although further device-level characterization is required to identify their contributions.

Charge retention was also evaluated by monitoring the stored voltage over time.
At 288~K, the voltage exhibited approximately exponential decay, with a fitted time constant of $\tau = 1.72 \pm 0.31$~s.
At cryogenic temperatures, no measurable voltage decay was observed within a 2-s measurement interval.
This result is consistent with reduced leakage at low temperatures, although the cryogenic retention time constant could not be determined within the measurement window.

\subsection{Room-Temperature Validation}
\label{subsec:room_temperature}

\begin{figure}[t]
\centering
\includegraphics[width=\columnwidth]{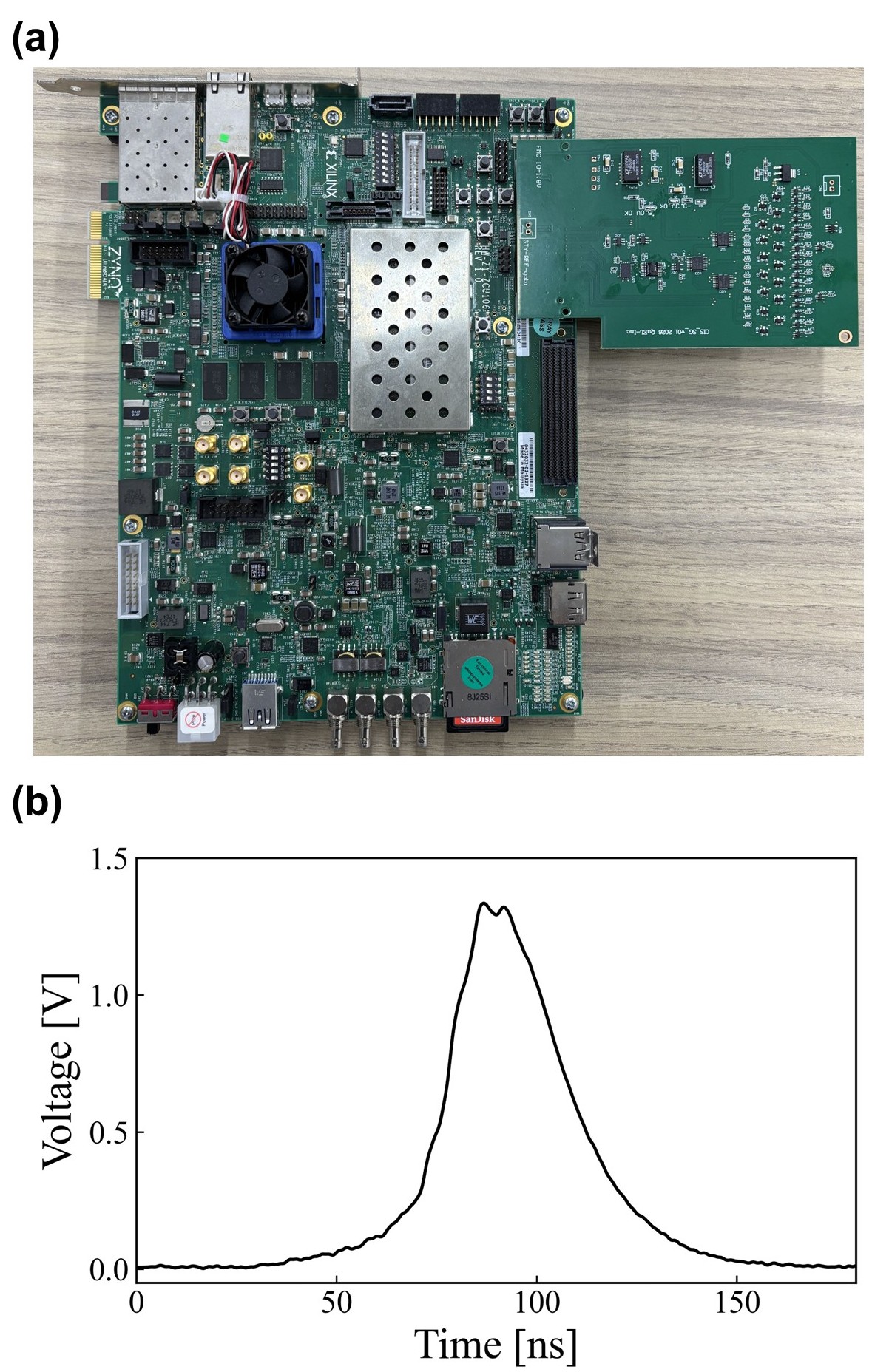}
\caption{
Room-temperature prototype and experimental result.
(a) Developed prototype.
(b) Gaussian-shaped waveform generated at a sampling rate of 160~Msps.
}
\label{fig:room_temperature}
\end{figure}

A complementary room-temperature prototype was developed to evaluate higher-speed waveform generation with improved signal routing.
Figs.~\ref{fig:room_temperature}(a) and (b) show the developed prototype and a representative Gaussian-shaped waveform, respectively.

The prototype employs a 12-bit DAC (AD5328) controlled by a ZCU106 FPGA platform through SPI.
The DAC output is amplified with a gain of two using an operational amplifier (TLV271) before being supplied to the demultiplexer network.
Waveform samples are programmed at 50~ksps and distributed to 16 analog memory cells through a two-stage demultiplexer network consisting of three TMUX6208 devices.
Each memory cell contains a 470-pF storage capacitor.
The stored voltages are sensed by MOSFET-based source followers (T2N7002AK) and sequentially selected through a readout multiplexer implemented using high-speed analog switches (SN74AUC1G66).

The room-temperature prototype follows the same basic architecture as the cryogenic prototype, comprising a low-speed waveform-programming path and a separate high-speed waveform-generation path.
However, the two implementations differ in the number of memory cells, the switching networks, and the signal-routing configuration.

In the room-temperature prototype, the digital control lines are physically separated from the analog output path.
As shown in Fig.~\ref{fig:room_temperature}(b), the prototype generated a Gaussian-shaped waveform at a sampling rate of 160~Msps without the pronounced switching-related spikes observed in the cryogenic experiment.

The achieved sampling rate of 160~Msps at room temperature is four times that of the 40~Msps cryogenic demonstration at 14~K.
This result demonstrates that the proposed waveform-generation architecture can operate at 160~Msps under room-temperature conditions, while its maximum achievable sampling rate at cryogenic temperatures remains to be determined.
The lower operating rate in the cryogenic prototype may be partly associated with the capacitive load introduced by the FPC interconnect, which can affect signal settling at high switching rates.
The improved waveform quality is also consistent with the separation of digital control and analog signal paths.
However, because the prototypes differ in both circuit configuration and operating conditions, the relative contributions of these factors have not yet been isolated.

\section{Conclusion}
\label{sec:conclusion}

We proposed an analog-memory-based arbitrary waveform generator that separates low-speed waveform programming from high-speed waveform generation.
By storing waveform samples as voltages in capacitor-based memory cells and accessing them through buffered switching networks, the proposed architecture enables repeated waveform generation without high-speed digital-to-analog conversion during waveform output.

A cryogenic prototype demonstrated Gaussian-shaped waveform generation at 14~K with a sampling rate of 40~Msps using a programming rate of 50~ksps.
No measurable voltage decay was observed over a 2-s interval at cryogenic temperatures.
A complementary room-temperature prototype achieved waveform generation at 160~Msps with reduced switching-related distortion.

These results demonstrate the feasibility of analog-memory-based waveform generation for cryogenic quantum-control applications.
By eliminating high-speed digital-to-analog conversion during waveform generation and limiting the number of switching transitions in the analog readout network, the proposed architecture offers the potential for reduced power consumption.
However, the total power consumption has not yet been quantitatively evaluated.
Future work will focus on evaluating power consumption,
improving waveform fidelity and cryogenic signal routing,
and integrating the waveform generator with a mixer and
local oscillator to generate microwave control signals
for superconducting qubits.

\clearpage

\bibliographystyle{IEEEtran}
\bibliography{ref}

\end{document}